\documentclass[a4paper,aps,pre,superscriptaddress,floatfix,nofootinbib,longbibliography,notitlepage,twocolumn]{revtex4-2}

\usepackage[utf8]{inputenc}
\usepackage[english]{babel}
\usepackage{amsmath,amssymb,amsfonts,amsthm,mathrsfs,amsopn}
\usepackage{mathtools}
\usepackage{graphicx}
\usepackage{xcolor}
\usepackage{bm}
\usepackage{array}
\usepackage{booktabs}
\usepackage{hyperref}

\def\be{\begin{equation}}
\def\ee{\end{equation}}

\begin{document}
\title{Rare but stable: hidden states in the 1D swarmalator model}

\author{Rommel Tchinda Djeudjo}
\affiliation{Department of Mathematics \& naXys, Namur Institute for Complex Systems, University of Namur, Rue Graf\'e 2, B-5000 Namur, Belgium}

\author{Kevin P. O'Keeffe}
\affiliation{Starling Research Institute, USA}

\author{Timoteo Carletti}
\email{timoteo.carletti@unamur.be}
\affiliation{Department of Mathematics \& naXys, Namur Institute for Complex Systems, University of Namur, Rue Graf\'e 2, B-5000 Namur, Belgium}

\begin{abstract}
The one-dimensional swarmalator model is usually understood through three
macroscopic states: synchrony, the phase wave, and asynchrony. We show that the
identical model also contains a much larger set of \emph{hidden} stable equilibria
-- families of $q$-twisted and $m$-clustered states -- which random initial
conditions almost never reach, because many are embedded in a degenerate neutral
manifold of vanishing global basin. They are nonetheless locally robust: under
Gaussian perturbations of per-coordinate amplitude $\varepsilon$ their survival
radii shrink only \emph{algebraically}, as $1/q$ and $1/m$, in sharp contrast to
the Gaussian basins and winding ceiling of twisted states on a nearest-neighbour
Kuramoto ring, both removed by mean-field coupling. A block-circulant Jacobian
fixes the stability in closed form, and far-kick experiments indicate the basins
are compact cores rather than tentacled sets. Several patterns reported in
\emph{extended} swarmalator models thus already exist, stable or neutrally stable,
in the minimal one.
\end{abstract}

\maketitle

\section{Introduction}

Swarmalators are mobile oscillators whose spatial motion and internal phase are
bidirectionally coupled, so that they synchronize in time while they self-assemble
in space~\cite{o2017oscillators}, building on earlier chemotactic-oscillator
models~\cite{tanaka2007general}. They model systems from Japanese tree
frogs~\cite{aihara2014spatio} and starfish embryos~\cite{tan2022odd} to magnetic
domain walls, active colloids and sperm
cells~\cite{hrabec2018velocity,yan2016reconfiguring,riedel2005self} and robotic
swarms~\cite{barcis2020sandsbots}. The one-dimensional variant, in which the
swarmalators run on a ring, is especially tractable: a change of variables decouples
it into a pair of Kuramoto-like models with three exact states (asynchronous, phase
wave, and synchronized) of known
stability~\cite{o2022collective,o2025stability,o2025global}. It has become a
benchmark to study how extra ingredients -- coupling disorder, pinning, forcing, phase
lag, finite range, time delay, inertia -- reshape collective
behaviour~\cite{yoon2022sync,o2022swarmalators,sar2023pinning,anwar2024forced,sar2025effects,okeeffe2026delay,okeeffe2026inertia,djeudjo2026role}.

The numerical results reported in such studies almost always start from random initial conditions; but which one of a
model stable states are actually \emph{reached} that way is a question of basin
stability~\cite{menck2013basin}, and the latter can be vanishingly small, preventing thus the system to achieve those states. For coupled
oscillators this is a classic theme -- the twisted states of a Kuramoto ring occupy
Gaussian basins~\cite{wiley2006size,groisman2025syncbasin}, and pulse-coupled
populations trap finite cluster fractions~\cite{okeeffe2026size}.

In this work we show that the swarmalator three classic states are only the most visible
members of a far larger family. Beside synchrony and phase wave, sit whole
families of $q$-twisted and $m$-clustered states, all linearly or neutrally
stable, an extensive number in all. They have gone unnoticed because random initial
conditions almost never reach them: they are rare but stable.

We give a complete analysis of the structured equilibria. The model gradient and
Hamiltonian structure allows us to solve every such state linear stability in closed
form, organized by the sign of one coupling combination $K=(J'+K')/2$, and to give a
precise completeness statement for the fixed-point problem off the degenerate
coupling axes. Most of the hidden states turn out to be \emph{neutrally} stable
rather than asymptotically attracting, which is exactly why they are hidden: they are
embedded in flat, high-dimensional neutral spaces, so their global basins are of
measure zero. What survives is a \emph{local} robustness, and its radius shrinks only
algebraically -- as $1/q$ and $1/m$ in the winding and cluster numbers. This is the
sharp opposite of twisted states on a Kuramoto ring~\cite{wiley2006size}, whose
basins are Gaussian, $e^{-kq^2}$, and which survive only below a winding ceiling;
mean-field coupling lifts both. The hidden states are not new physics requiring new
interactions -- they are the generic, exactly-describable furniture of the minimal
model.

% =====================================================================
% Body. HOUSE NOTATION: (x,theta)/(J',K'); xi=x+theta, eta=x-theta /(J,K),
% K=(J'+K')/2, J=(J'-K')/2.  Structure -> catalog+completeness -> phase diagram
% -> LOCAL ROBUSTNESS (climax) -> geometry -> discussion.
% =====================================================================

\section{Model and gradient structure}

We study the identical 1D swarmalator model~\cite{o2022collective},
\begin{align}
  \dot x_i &= \frac{J'}{N}\sum_{j}\sin(x_j-x_i)\cos(\theta_j-\theta_i),
  \label{eq:modelx}\\
  \dot\theta_i &= \frac{K'}{N}\sum_{j}\sin(\theta_j-\theta_i)\cos(x_j-x_i),
  \label{eq:modeltheta}
\end{align}
with $x_i,\theta_i\in\mathbb S^1$, identical units ($v_i=\omega_i=0$), and bare
spatial and phase couplings $J',K'$.

\emph{Rainbow variables.} The change of variables $\xi=x+\theta$, $\eta=x-\theta$
recasts the model as two Kuramoto models coupled through the rainbow order parameters
$re^{i\phi}=\langle e^{i\xi}\rangle$, $se^{i\psi}=\langle e^{i\eta}\rangle$,
\begin{align}
  \dot\xi_i &= K\,r\sin(\phi-\xi_i) + J\,s\sin(\psi-\eta_i),\label{eq:xidot}\\
  \dot\eta_i &= J\,r\sin(\phi-\xi_i) + K\,s\sin(\psi-\eta_i),
\end{align}
where the \emph{rainbow couplings}
\begin{equation}
  K=\tfrac12(J'+K'),\qquad J=\tfrac12(J'-K')
  \label{eq:JKdef}
\end{equation}
combine the bare couplings $(J',K')$. We use unprimed $(J,K)$ for the rainbow pair
throughout, and primed $(J',K')$ for the bare spatial/phase pair of
Eqs.~\eqref{eq:modelx}--\eqref{eq:modeltheta}; in particular ``$K>0$'' always means
$J'+K'>0$, not $K'>0$.

\emph{Gradient and Hamiltonian structure.} In general the flow has a Helmholtz
decomposition~\cite{o2025global}, $\dot z=-\nabla V+\mathsf S\nabla H$, where
$\mathsf S=\big(\begin{smallmatrix}0&1\\-1&0\end{smallmatrix}\big)$ rotates each
$(x_i,\theta_i)$ pair and the mean-field potentials are
\begin{equation}
  V=-\tfrac{N}{4}K(r^2+s^2),\qquad H=\tfrac{N}{4}J(r^2-s^2).
  \label{eq:VH}
\end{equation}
The relaxational (gradient) part $V$ is set by $K$, the conservative (Hamiltonian)
part $H$ by $J$.

\emph{Transformed gradient structure.} The dynamics also follow the single potential
\begin{equation}
\begin{split}
  U&=-\frac{1}{2N}\sum_{i,j}\cos(x_i{-}x_j)\cos(\theta_i{-}\theta_j)\\
   &=-\tfrac{N}{4}(r^2+s^2),
\end{split}
  \label{eq:U}
\end{equation}
as $\dot x_i=-J'\partial_{x_i}U$, $\dot\theta_i=-K'\partial_{\theta_i}U$, with metric
$\mathrm{diag}(J',\dots,K',\dots)$. When $J',K'>0$ this metric is positive definite
and the flow is a transformed gradient~\cite{pedergnana2022exact,o2025global} that
\emph{descends} $U$; when $J',K'<0$ it \emph{ascends} $U$ (a transformed gradient of
$-U$). In either same-sign quadrant $\dot U=-\sum_i[J'(\partial_{x_i}U)^2+
K'(\partial_{\theta_i}U)^2]$ is sign-definite, so $\pm U$ is a Lyapunov function and
the flow is purely relaxational. For opposite signs the metric is indefinite, the
Hamiltonian part is essential, and on the line $K=0$ the flow is purely Hamiltonian
with $H$ conserved. Either way $\operatorname{sgn}(K)$ organizes the stability
landscape, and the degeneracy of $U$ in Eq.~\eqref{eq:U} is what makes most of the
model states hidden.

\section{Exact stability catalog of structured equilibria}
\label{sec:catalog}

\emph{Closed-form stability.} Differentiating Eq.~\eqref{eq:xidot}, the Jacobian
about \emph{any} configuration is, in $(\xi,\eta)$ block form,
\begin{equation}
  M=\begin{pmatrix} K\,L^\xi & J\,L^\eta\\[2pt]
                    J\,L^\xi & K\,L^\eta \end{pmatrix},
  \label{eq:Mblock}
\end{equation}
where $L^\phi$ is the Kuramoto Laplacian of the configuration $\phi^*$,
$L^\phi_{ij}=\tfrac1N\cos(\phi^*_j-\phi^*_i)$ ($i\ne j$),
$L^\phi_{ii}=-\tfrac1N\sum_{l\ne i}\cos(\phi^*_l-\phi^*_i)$. For the twisted and
clustered states below $L^\xi,L^\eta$ are circulant, so $M$ block-diagonalizes into
$2\times2$ blocks, one per Fourier mode $k$,
\begin{align}
  M_k&=\begin{pmatrix} K\mu^\xi_k & J\mu^\eta_k\\
                      J\mu^\xi_k & K\mu^\eta_k\end{pmatrix},\\
  \lambda_\pm&=\tfrac12 K(\mu^\xi_k+\mu^\eta_k)\nonumber\\
  &\quad\pm\tfrac12\sqrt{K^2(\mu^\xi_k-\mu^\eta_k)^2+4J^2\mu^\xi_k\mu^\eta_k},
  \label{eq:Mk}
\end{align}
with $\mu_k=\tfrac12$ at $k\equiv\pm w$ for a splay of winding $w$ and $\mu_k=-1$
($k\ne0$) for a synchronized coordinate. We have verified that
Eqs.~\eqref{eq:Mblock}--\eqref{eq:Mk} reproduce the full $2N\times2N$ numerical
spectrum of every state below to machine precision (Appendix~\ref{app:numerics}).

\begin{figure}[!ht]
  \includegraphics[width=\linewidth]{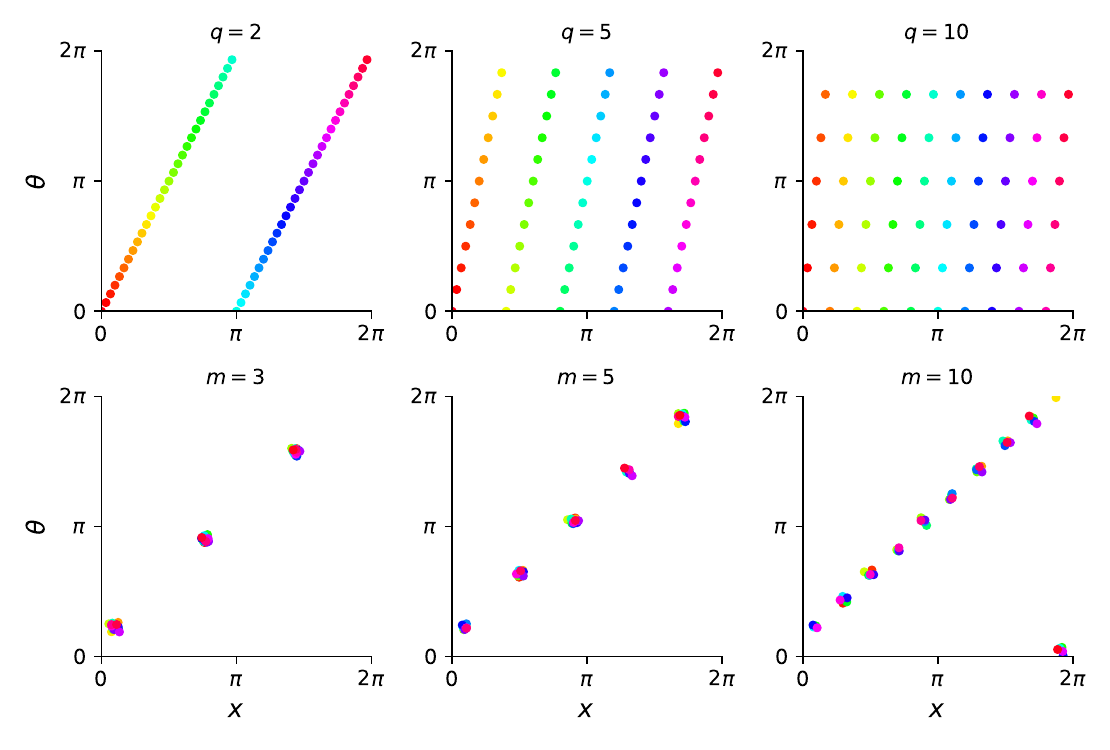}
  \caption{The two hidden families ($N=60$). Top: $q$-twists for $q=2,5,10$ -- the
  swarmalators lie on $q$ winding bands of the $(x,\theta)$ torus. Bottom:
  diagonal $m$-clusters for $m=3,5,10$ -- $m$ correlated dots on the diagonal. All
  shown are linearly (Lyapunov) stable -- twists for $K<0$, clusters for
  $J'+K'>0,\ J'K'<0$ -- with no upper bound on $q$ or $m$, in contrast to the
  Kuramoto ring. The degenerate small cluster numbers $m=2$ (synchrony) and $m=4$
  are treated in Appendix~\ref{app:smallm}.}
  \label{fig:qgallery}
\end{figure}

\emph{The twisted family.} The $(a,b)$-twist $x_p=ap\Delta+c_1$,
$\theta_p=bp\Delta+c_2$ ($\Delta=2\pi/N$) has $\xi,\eta$ winding $a{+}b,\,a{-}b$.
(i) In the case, $a,b\ne0$, $a\ne\pm b$ ($r=s=0$): a single distinct nonzero eigenvalue
$\lambda=K/2$ of multiplicity four, and $2N-4$ zero modes do exist; they are stable if and only if $K<0$ -- the
$q$-twist family $(1,q)$ [Fig.~\ref{fig:qgallery}, top]. For a given $N$ the distinct
hidden twists run over $q=2,\dots,\lceil N/2\rceil-1$ modulo aliasing: $q\equiv0,\pm1$
reduce to the catalogued degeneracies below, and for even $N$ the endpoint $q=N/2$ is
excluded -- there $\theta_p=\pi p$ collapses to an antipodal pair, the two active
harmonics alias ($1{+}q\equiv1{-}q\bmod N$), and the spectrum degenerates to that of
spatial/phase sync ($J'/2$ and $K'/2$, each twice), stable only for $J',K'<0$. There
is one harmless finite-$N$ endpoint degeneracy: for even $N$, $q=N/2-1$ makes one
rainbow winding self-conjugate, giving $K$ (once) and $K/2$ (twice) instead of
$K/2$ (four times), but the stability condition remains $K<0$. Apart from this
aliasing the eigenvalue is $q$-independent, so \emph{every} retained winding is stable
-- there is no Kuramoto-ring $N/4$ ceiling (Sec.~\ref{sec:basins}).
(ii) In the case, $a=0$ or $b=0$
($r=s=0$): nonzero eigenvalues are $J'/2$ (mult.\ two), $K'/2$ (mult.\ two) and $2N-4$
zero modes; they are stable if and only if $J',K'<0$ -- spatial synchrony $(0,1)$ and phase synchrony
$(1,0)$. 

(iii) Finally, if $a=\pm b$: one rainbow order parameter is unity -- the phase wave.

\emph{The clustered family.} The diagonal $m$-cluster states $x_k=c_1+\alpha_k$,
$\theta_k=c_2+\alpha_k$, $\alpha_k=\tfrac{2\pi}{m}(k\bmod m)$ ($N$ a multiple of
$m$), have $\eta_k=c_1-c_2$ constant, hence $s=1$, while $\xi_k=2\alpha_k+\text{const}$
splays into clusters [Fig.~\ref{fig:qgallery}, bottom]. For $m\ge3$ the within-cluster
modes are neutral in $\xi$ and decay in $\eta$ ($-K$, multiplicity $N-m$), requiring
$K>0$; the cluster-level modes are circulant -- the $m$ clusters wind twice in $\xi$,
so the active harmonic $k\equiv\pm2$ reduces to Eq.~\eqref{eq:Mk} with
$\mathrm{tr}(M_k)=-K/2$, $\det (M_k)=-\tfrac12J'K'$ (the four \emph{block} modes), while the
remaining cluster harmonics $k\not\equiv0,\pm2\pmod m$ have $\mu^\xi_k=0$,
$\mu^\eta_k=-1$ (the clusters are $\eta$-synchronized) and each contribute one $-K$
mode and one zero. The decay eigenvalue $-K$ therefore has total multiplicity
$(N-m)+(m-3)=N-3$ and there are $N-1$ zero modes. Both block roots have negative real
part precisely when $K>0$ and $J'K'<0$: this real part is $-K/4$ in the oscillatory
regime that fills most of the wedge, and the roots become real but still negative
in the thin sliver $|K'|/J'\lesssim0.03$ adjoining the axis. The generic family
($m=3,5,6,\dots$)
is therefore Lyapunov stable for $J'+K'>0,\ J'K'<0$ -- the phase-wave region
$-J'<K'<0$ of Ref.~\cite{o2022collective}; we have confirmed this to $m=12$. These
$m$-clusters share the phase wave order parameters $(0,1)$ and, with these
multiplicities, its \emph{entire} spectrum (Table~\ref{tab:catalog}): they are the
discrete $m$-fold microstates of the phase wave, coarse-grainings of the same $s=1$
splay. Two small cluster numbers are exceptional and are treated
separately in Appendix~\ref{app:smallm}: $m=2$ collapses to $r=s=1$ (it is the
\emph{synchronized} state, stable only for $J',K'>0$, and is \emph{not} a member of
the cluster family), and $m=4$ has a self-conjugate active harmonic ($+2\equiv-2
\bmod 4$) whose collapsed block yields a marginal oscillatory pair
$\lambda=\pm i\sqrt{-J'K'}$ (purely imaginary, hence neutral but \emph{not} zero) --
still no positive eigenvalue, so $m=4$ remains Lyapunov stable in the same wedge, but is
even more degenerate than the generic case.

\begin{table*}[!t]
\caption{Structured equilibria of the identical 1D swarmalator and their linear
stability from Eq.~\eqref{eq:Mk}; $K=\tfrac12(J'+K')$, $J=\tfrac12(J'-K')$.
Multiplicities are for the indicated $(r,s)$ family; ``neutral'' modes are the
$N$-dependent zero eigenvalues. ``Stability type'' distinguishes asymptotically
stable (a.s.) from Lyapunov / neutrally stable (n.s.). The reflection
$\theta\to-\theta$ swaps $(r,s)\to(s,r)$, so each row implies its mirror (e.g.\ the
phase wave $(0,1)$ and its partner $(1,0)$). Sync and the phase wave are the
generic, full-basin outcomes (Sec.~\ref{sec:phase}); the remaining rows are the
hidden states. ``Block'' denotes the four cluster-level circulant modes, of negative
real part throughout the wedge ($-K/4$ in the oscillatory regime). For the async row,
$K<0$ means the manifold $\mathcal M$ is globally attracting, with each point
neutrally stable \emph{along} $\mathcal M$. The Hamiltonian line $K=0$ is excluded
(Sec.~\ref{sec:phase}).}
\label{tab:catalog}
\footnotesize
\begin{ruledtabular}
\begin{tabular}{lcccc}
state & $(r,s)$ & nonzero eig.\ (mult.) & \# zero & type / stable iff\\
\hline
sync ($m{=}2$) & $(1,1)$ & $-J'(N{-}1),\,-K'(N{-}1)$ & $2$ & a.s.; $J',K'>0$\\
phase wave & $(0,1)$ & $-K(N{-}3);\ \text{block}(4)$ & $N{-}1$ & n.s.; $K>0,J'K'<0$\\
$m$-cluster ($m\!\ge\!3$) & $(0,1)$ & $-K(N{-}3);\ \text{block}(4)$ & $N{-}1$ & n.s.; $K>0,J'K'<0$\\
\,$m{=}4$ (degenerate) & $(0,1)$ & $-K(N{-}2);\ \pm i\sqrt{-J'K'}$ & $N$ & n.s.; $K>0,J'K'<0$\\
spatial/phase sync & $(0,0)$ & $\tfrac{J'}{2}(2),\,\tfrac{K'}{2}(2)$ & $2N{-}4$ & n.s.; $J',K'<0$\\
$q$-twist (generic) & $(0,0)$ & $\tfrac{K}{2}(4)$ & $2N{-}4$ & n.s.; $K<0$\\
async (disordered) & $(0,0)$ & $0$ & $2N{-}4$ & n.s.\ on $\mathcal M$; $K<0$\\
\end{tabular}
\end{ruledtabular}
\end{table*}

\emph{A completeness statement.} Table~\ref{tab:catalog} exhausts the structured
equilibria in the following precise sense. \emph{Consider the identical model
Eqs.~\eqref{eq:modelx}--\eqref{eq:modeltheta} at finite $N$, off the degenerate
coupling axes $J'K'=0$ and off the Hamiltonian line $K=0$, and ask for linearly
stable fixed points modulo the global rotations and the reflection
$\theta\to-\theta$.} The disordered manifold $\mathcal M$ (the async row) carries a
continuum of $r=s=0$ configurations -- the $q$-twists and splay-synchronized states
are its symmetric, circulant points -- so it remains to rule out \emph{uncatalogued
mixed} fixed points with $0<r,s<1$ (the catalogued sync state $r=s=1$ aside). At such
a fixed point the force balance
$\big(\begin{smallmatrix}Kr&Js\\Jr&Ks\end{smallmatrix}\big)
\big(\begin{smallmatrix}\sin(\phi-\xi_i)\\ \sin(\psi-\eta_i)\end{smallmatrix}\big)=0$
has determinant $rs(K^2-J^2)=rs\,J'K'$. Off the axes ($J'K'\ne0$) this is nonzero,
forcing $\xi_i\in\{\phi,\phi+\pi\}$, $\eta_i\in\{\psi,\psi+\pi\}$: every mixed fixed
point is a four-cluster state. Its within-cluster (zero-sum) modes give the block
$B=-\big(\begin{smallmatrix}K\varepsilon_\xi r&J\varepsilon_\eta s\\
J\varepsilon_\xi r&K\varepsilon_\eta s\end{smallmatrix}\big)$ with
$\det (B)=\varepsilon_\xi\varepsilon_\eta\,rs\,J'K'$; since $\varepsilon_\xi
\varepsilon_\eta=+1$ for two clusters and $-1$ for the other two, one pair has
$\det B<0$ and hence a positive eigenvalue. \emph{Every off-axis mixed fixed point is
linearly unstable}, so within this class no mixed attractor exists, and the stable
families reduce to those in Table~\ref{tab:catalog} (together with the continuum
$\mathcal M$ for $K<0$). The degenerate axes $J'K'=0$ are the boundary $K^2=J^2$ where
$\det (B)$ vanishes; an unoccupied unstable pair there reduces the state to a two- or
three-cluster, itself unstable. Let us emphasize what this statement does and does not
cover: it is a linear-stability classification of \emph{fixed points} of the
relaxational dynamics, away from the axes and away from $K=0$; it does not claim to
enumerate invariant sets of the conservative flow on $K=0$, which we exclude
deliberately. Within its scope the model is \emph{extensively} multistable: for
$K<0$ alone it has $\sim N$ distinct twisted and splay states.

\section{Phase diagram and regimes}
\label{sec:phase}

\begin{figure}[!ht]
  \includegraphics[width=0.8\linewidth]{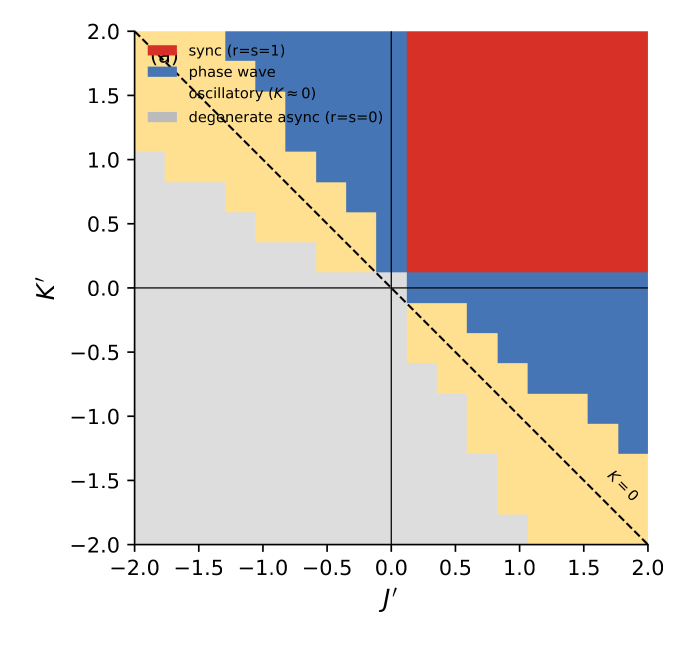}\\
    \includegraphics[width=0.8\linewidth]{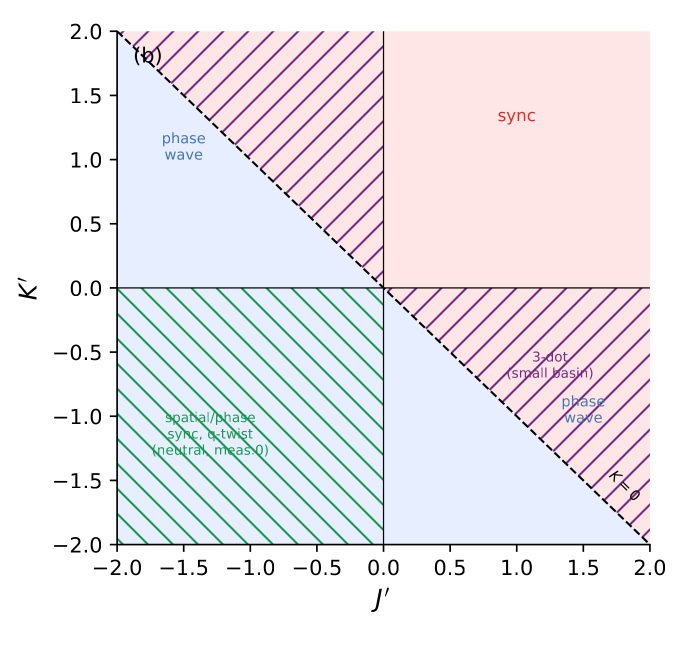}
  \caption{(a) Generic outcome from random initial conditions over $(J',K')$
  ($N=16$): sync, phase wave, the Hamiltonian line $K=0$ (dashed; $K'=-J'$, apparent
  width is grid resolution), and the degenerate asynchronous manifold $\mathcal M$
  for $K<0$. (b) Analytic stability regions of the structured states
  (Table~\ref{tab:catalog}), organized by $\operatorname{sgn}K$.}
  \label{fig:phase}
\end{figure}

Because $U$ depends on the microstate only through $(r,s)$, the relaxational flow
drives $r^2+s^2$ according to $\operatorname{sgn}(K)$, giving three regimes
[Fig.~\ref{fig:phase}].

\emph{$K>0$: relaxational order.} $U$ is minimized by maximizing $r^2+s^2$; random
initial conditions reach an ordered state with full basin -- global sync for
$J',K'>0$, or the phase wave $(0,1)$ for opposite-sign couplings with $J'+K'>0$.

\emph{$K=0$ ($K'=-J'$): a Hamiltonian line, excluded from the classification.} The
gradient part vanishes; the flow conserves $H$ [Eq.~\eqref{eq:VH}] (relative drift
$\sim10^{-9}$ over the run) and the order parameters oscillate without decay, with an amplitude
$\sim N^{-1/2}$ from random initial conditions. This is a singular boundary of the
relaxational picture: the flow is conservative there, so the equilibrium language
used off the line does not apply, and a conservative flow may carry periodic,
quasiperiodic, or more intricate invariant sets. We therefore treat $K=0$ only as a
marginal boundary -- the line on which, fluctuations cease to damp -- and leave its
invariant-set structure to future work. In Fig.~\ref{fig:phase}(a) it is drawn as a
single dashed line $K'=-J'$; any apparent width is the $(J',K')$ grid resolution.

\emph{$K<0$: a degenerate disordered manifold.} $U$ is maximized by minimizing
$r^2+s^2$; every random initial condition relaxes to exactly $r=s=0$, the manifold
\begin{equation}
  \mathcal M=\{(x,\theta):r=s=0\},\qquad \dim\mathcal M=2N-4,
  \label{eq:manifold}
\end{equation}
on which $U$ is flat. The eigenvalues of Eq.~\eqref{eq:Mk} carry no $N$, so every
boundary in Table~\ref{tab:catalog} is $N$-independent and the phase diagram has a
well-defined $N\to\infty$ limit (we confirm random initial conditions still collapse
to $r=s=0$ at $N=192$). The relaxational picture is rigorous where the metric is
definite -- the same-sign quadrants, $J',K'>0$ being the global synchronization
theorem of Ref.~\cite{o2025global}. For opposite-sign couplings the gradient
structure is lost~\cite{o2025global}; there Table~\ref{tab:catalog} leaves only the
ordered state or $\mathcal M$ stable, and trajectories are seen to reach them with no
limit cycles off $K=0$, so global attraction is established numerically while local
stability and the completeness statement are exact.

\section{Local robustness of the hidden states}
\label{sec:basins}

The model is extensively multistable, yet from random initial conditions one sees
only sync, phase wave, or a disordered point of $\mathcal M$. The structured
states -- spatial and phase synchrony, the $q$-twists and $m$-clusters
[Fig.~\ref{fig:qgallery}] -- are all linearly stable, but a $10^6$-sample search at
each $N$ finds none of them. We must therefore distinguish two different notions of
basin size.

\emph{Global basin stability versus local survival radius.} The \emph{global basin
stability}~\cite{menck2013basin} of a hidden state -- the probability that a
uniformly random $(x_i,\theta_i)\sim\mathrm{Unif}(\mathbb T^{2N})$ reaches it -- is
zero or numerically invisible in our simulations ($<10^{-6}$ at every $N$). This is
expected for states embedded in the degenerate manifold $\mathcal M$, or coexisting
with the dominant phase wave. We therefore measure a different, local object. Starting
\emph{from} the exact state we apply independent Gaussian kicks
$x_p\mapsto x_p+\varepsilon\xi_p$, $\theta_p\mapsto\theta_p+\varepsilon\zeta_p$, with
$\xi_p,\zeta_p$ standard normals, we numerically integrate the system, and ask whether the trajectory returns to
the same labeled state. The resulting threshold $\varepsilon^\ast$ is the radius of
the local \emph{basin core} in this perturbation ensemble, not a global basin volume.
Here $\varepsilon$ is the per-coordinate standard deviation of the kick; the
corresponding Euclidean displacement in the $2N$-dimensional phase space is typically
$\|\delta z\|_2\sim\varepsilon\sqrt{2N}$, so ``radius'' below always means this
per-coordinate amplitude.

\begin{figure*}[!t]
  \includegraphics[width=0.86\textwidth]{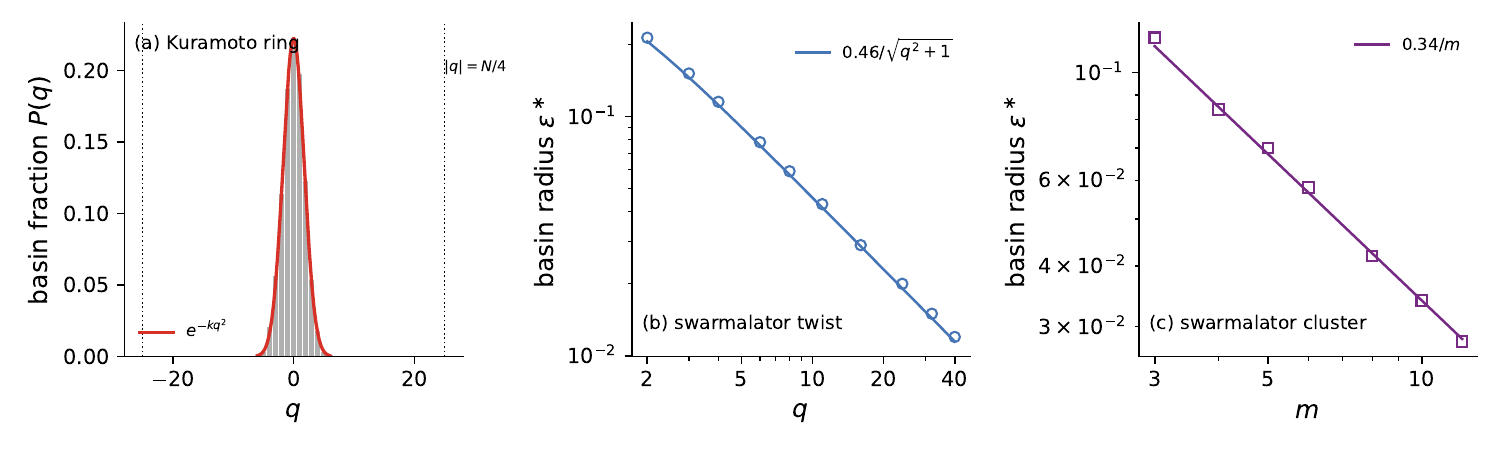}
  \caption{Local survival radius versus winding or cluster number -- the mean-field
  opposite of the ring. (a) On a nearest-neighbour Kuramoto ring, twisted states have
  finite, Gaussian global basins $P(q)\sim e^{-kq^2}$ and exist only for $|q|<N/4$
  (dotted). (b,c) In the mean-field swarmalator the twists and clusters are neutral --
  their \emph{global} basin under random initial conditions is of measure zero
  (numerically $<10^{-6}$), but the \emph{local} survival radius under Gaussian kicks
  is algebraic, $\varepsilon^\ast_q=0.46/\sqrt{q^2+1}$ ($\to0.46/q$ at large $q$) and
  $\varepsilon^\ast_m=0.34/m$, with no ceiling. Scaling exponents are exact; the
  prefactor is the empirical threshold of Eq.~\eqref{eq:epsstar}.}
  \label{fig:contrast}
\end{figure*}

\emph{A geometric scaling law.} Both kinds of state are robust-but-narrow for the
same reason: a high-dimensional \emph{neutral} space makes their labeling order
parameter quasi-conserved. A $q$-twist has the single distinct nonzero eigenvalue
$K/2$ and $2N-4$ zero modes; an $m$-cluster is neutral in $\xi$ within each cluster. A
Gaussian kick is therefore almost entirely neutral, and disturbs the labeling order
parameter -- $w_q=|\langle e^{i(qx-\theta)}\rangle|$ for a twist,
$X_m=|\langle e^{imx}\rangle|$ for a cluster -- only weakly. On the exact state the
labeling phase is site-independent ($qx_p-\theta_p$ and $mx_p$ are constant); the kick
shifts it by $\varepsilon(q\xi_p-\zeta_p)$ for a twist, by $m\varepsilon\xi_p$ for a
cluster -- Gaussians of variance $\varepsilon^2(q^2+1)$ and $m^2\varepsilon^2$.
Averaging the phase factor over the population gives its Gaussian characteristic
function, so the labeling order parameter relaxed onto the neutral space -- written
$\bar w_q,\bar X_m$, the value left once the transverse mode has decayed back -- is
set by the kick geometry alone,
\begin{equation}
  \bar w_q=e^{-\frac12\varepsilon^2(q^2+1)}+O(N^{-1/2}),\quad
  \bar X_m=e^{-\frac12 m^2\varepsilon^2}+O(N^{-1/2}),
  \label{eq:scramble}
\end{equation}
the labeling phase amplifying the kick by $q$ (or $m$); the $O(N^{-1/2})$ term is the
finite-$N$ sampling fluctuation of the order parameter, which sets a floor at large
$q,m$ or small $N$. The state survives while this stays above an $O(1)$ threshold;
setting $\bar w_q=w_c$, $\bar X_m=X_c$ gives the algebraic, $N$-independent survival radii
\begin{equation}
  \varepsilon^\ast_q=\frac{\sqrt{-2\ln w_c}}{\sqrt{q^2+1}}\ \ (\text{twists}),\qquad
  \varepsilon^\ast_m=\frac{\sqrt{-2\ln X_c}}{m}\ \ (\text{clusters}).
  \label{eq:epsstar}
\end{equation}
The exponents -- the $1/\sqrt{q^2+1}$ and $1/m$ scaling, and the data collapse below
-- are \emph{exact geometric consequences} of the neutral structure and carry no free
parameter. The \emph{prefactor}, by contrast, is empirical: it is fixed by the
threshold $w_c$ (resp.\ $X_c$) at which the neutral approximation yields to the
transverse mode. We measure $w_c\approx0.90$ for twists, so
$\varepsilon^\ast_q\to0.46/q$ at large $q$, and a somewhat higher $X_c\approx0.94$ for
clusters, giving $\varepsilon^\ast_m\approx0.34/m$. We confirm both laws over a decade
in $q$ and $m$ and independent of $N$ [Fig.~\ref{fig:contrast}(b,c)]; the relaxed order
parameter collapses onto the single variable $\varepsilon\sqrt{q^2+1}$ (resp.\
$\varepsilon m$), the signature of the geometric mechanism. We therefore claim the
\emph{scaling} as exact and the \emph{prefactor} as an empirically determined
threshold, not an exact constant.

\emph{Mean-field versus ring.} This is the sharp counterpart of twisted states on a
\emph{nearest-neighbour} Kuramoto ring~\cite{wiley2006size}. There, local coupling
makes the winding a topological sector with an extensive barrier: only $|q|\lesssim
N/4$ are stable and their basins are Gaussian, $e^{-kq^2}$
[Fig.~\ref{fig:contrast}(a)] -- a scaling conjectured in 2006~\cite{wiley2006size},
supported analytically~\cite{groisman2025syncbasin} and recently proved for the
octopus geometry~\cite{groisman2026tentacles}. The mean-field coupling here does the
opposite on both counts -- \emph{every} retained winding is stable, and the local
survival radius is the far milder algebraic $1/q$. The async manifold itself is the
two-coordinate version of incoherence in the repulsive Kuramoto
model~\cite{strogatz1991stability,hong2011kuramoto} (a $q$-splay there is neutral with
transverse eigenvalue $g/2$, our $K/2$ with $g\to K$); what is genuinely new is the
$m$-cluster branch on $\{s=1\}$, with no analogue in a single Kuramoto model.

\section{Basin geometry: boxy or octopus?}

A local survival radius does not determine the global shape of a basin. In
high-dimensional systems a basin may have a small robust core near the state plus
long, thin tentacles that hold almost all of its volume and reach far across phase
space; Zhang and Strogatz~\cite{zhang2021tentacles} showed that systems with
subexponentially many attractors -- the Kuramoto ring among them -- are generically
\emph{octopus-like} in just this way. Establishing the $1/q,1/m$ \emph{core} radius of
Sec.~\ref{sec:basins} therefore says nothing yet about tentacles; that requires a
separate far-reach test.

\begin{figure*}[!t]
  \includegraphics[width=0.86\textwidth]{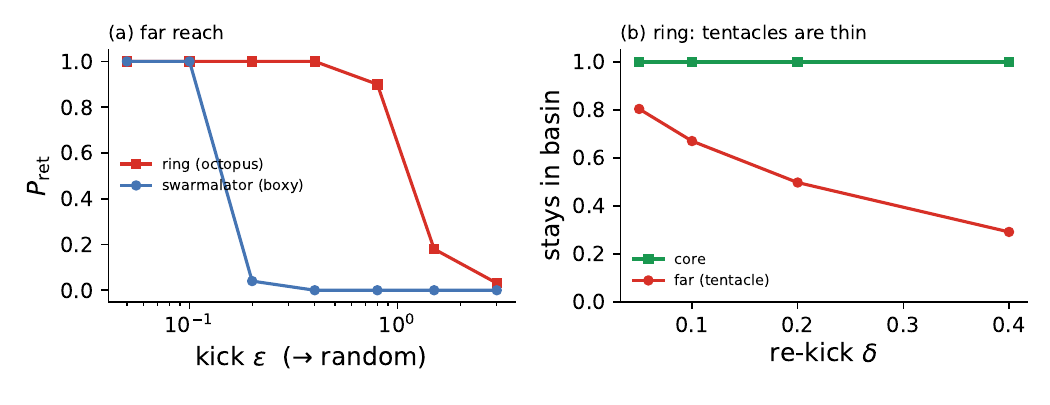}
  \caption{(a) Return probability from a kick of per-coordinate amplitude
  $\varepsilon$, out to the random-initial-condition scale: the ring twist retains a
  finite far reach (tentacles), the swarmalator $m$-cluster ends at its core (boxy);
  the $q$-twist behaves identically. (b) The ring's far-returning points are fragile
  under a re-kick (thin tentacles), its core robust.}
  \label{fig:geom}
\end{figure*}

We make the logic explicit. The local kicks of Sec.~\ref{sec:basins} measure the core
radius $\varepsilon^\ast$. To test for tentacles we instead grow the kick to the
random-initial-condition scale and measure the far-return probability
$P_{\rm ret}(\varepsilon)$ [Fig.~\ref{fig:geom}]: a nonzero $P_{\rm ret}$ at large
$\varepsilon$ would signal octopus-like reach, whereas $P_{\rm ret}\to0$ means the
basin ends at its core. On the ring, random initial conditions reach a given winding
with finite probability, and those distant points are fragile -- a small re-kick flips
half of them to another winding [Fig.~\ref{fig:geom}(b)] -- the signature of thin
tentacles. In the swarmalator $P_{\rm ret}\to0$ for both twists and clusters: past
$\varepsilon^\ast$ the basin simply ends, with no far reach. The hidden-state basin is
a compact core and nothing more.

The reason is locality: a tentacle forms when a distant configuration drains into a
state through local rearrangements that interleave neighbouring basins, which the
nearest-neighbour ring permits and the mean-field model does not. Each hidden-state
basin is therefore an isolated core in a sea that flows to $\mathcal M$ or the phase
wave -- the geometric counterpart of its algebraic, ceiling-free core radius
(Sec.~\ref{sec:basins}): both follow from mean-field coupling erasing the spatial
structure the ring exploits.

\section{Discussion}

We have solved the structured-equilibrium problem of identical 1D swarmalator: a
single parameter $K=(J'+K')/2$ organizes the landscape, every structured state stability follows in closed form from one block-circulant Jacobian, the fixed-point
catalog is complete off the degenerate axes, and the local robustness of the many
hidden states obeys an exact algebraic scaling law. The picture is unified by
neutrality -- a degenerate manifold of macrostates and high-dimensional neutral spaces
within them -- which makes the hidden states both order-parameter blind and
geometrically, rather than topologically, protected. The fact that the same model is
\emph{algebraic} where the Kuramoto ring is \emph{Gaussian} pinpoints what mean-field
coupling does to a twisted state: it removes the stability ceiling and softens the
local basin. The correct one-line summary is not that these states have large basins
-- globally they have none -- but that they are exact stable or neutrally stable
equilibria with compact local robustness cores whose radii decay only algebraically.

The practical lesson concerns how collective states are discovered. Several of the
patterns catalogued here -- phase and spatial synchrony, the three-dot, the
$q$-twists -- have counterparts that have been reported as new features of
\emph{extended} swarmalator models~\cite{sar2025effects}. We find that stable or
neutrally stable versions are already present in the minimal
model~\cite{o2022collective}, merely invisible to the random initial conditions used
to explore it. The role of the added ingredients may therefore be not only to create
such patterns from scratch but also to enlarge their basins, lift their neutral
degeneracy, or make them reachable from generic initial conditions -- and an
extended-model state and its minimal-model counterpart need not be the same dynamical
object. Before attributing a regime to a new coupling or external drive, it is worth
checking whether it is already a stable -- or neutral -- solution of the bare model,
and characterizing its basin rather than its mere existence.

Two directions follow naturally. Heterogeneity ($v_i,\omega_i\ne0$) will lift the
neutral manifold, turning the exactly-degenerate hidden states into long-lived
transients whose lifetime should inherit the $1/q,1/m$ scaling; and the Hamiltonian
line $K=0$, where order parameters oscillate without damping, invites a dedicated
study of its conservative invariant sets.

\appendix

\section{Degenerate small cluster numbers}
\label{app:smallm}

The generic clustered-family argument of Sec.~\ref{sec:catalog} assumes the cluster
configuration $\xi_k=2\alpha_k+\text{const}$ genuinely splays into $m$ distinct
clusters winding twice, with distinct active harmonics $k\equiv\pm2$. Two small $m$
violate this and must be handled separately. We verified all statements below by
forming the exact $2N\times2N$ Jacobian and comparing its spectrum to the block
formula Eq.~\eqref{eq:Mk}, for $m=2,\dots,12$ with $N$ a multiple of $m$ (agreement
to $\sim10^{-10}$).

\emph{$m=2$ is the synchronized state.} For $m=2$, $\alpha_k\in\{0,\pi\}$ so
$2\alpha_k\in\{0,2\pi\}\equiv0$: the $\xi$ coordinate is \emph{synchronized}, not
split, and $\eta$ is constant, giving $r=s=1$. The ``two-cluster diagonal state'' is
thus identical to global synchrony, with spectrum $-J'$ (mult.\ $N-1$), $-K'$ (mult.\
$N-1$) and two global-rotation zero modes. It is stable in the sync quadrant
$J',K'>0$ and \emph{unstable} in the cluster wedge $J'+K'>0,\ J'K'<0$ (at
$(J',K')=(2,-0.5)$ its largest eigenvalue is $+\tfrac12$). It therefore belongs to the
sync row of Table~\ref{tab:catalog}, not to the $m$-cluster family, and is excluded
from the clustered-state figure and stability claim.

\emph{$m=4$ has a self-conjugate active harmonic.} For generic $m\ge3$ the cluster
configuration winds twice, so the active circulant harmonics $k\equiv\pm2$ are
distinct and their $2\times2$ block has roots with real part $-K/4<0$ in the wedge --
those cluster-level modes \emph{decay} ($-K/4$ real part). For $m=4$, $+2\equiv-2\bmod4$:
the active harmonic is self-conjugate, the two harmonics coincide, and the collapsed
block yields a single \emph{marginal oscillatory} pair $\lambda=\pm i\sqrt{-J'K'}$ --
purely imaginary, hence neutral but \emph{not} zero. The remaining cluster harmonics
$k=1,3$ still give one $-K$ mode and one zero each, so the $m=4$ spectrum is $-K$
(mult.\ $N-2$), the pair $\pm i\sqrt{-J'K'}$, and $N$ zero modes -- versus the generic
$-K$ (mult.\ $N-3$), block$(4)$, and $N-1$ zeros. Crucially, no eigenvalue acquires a
positive real part: $m=4$ remains Lyapunov stable in the same wedge $J'+K'>0,\ J'K'<0$,
but is more degenerate than the generic case, with the cluster-level restoring block
replaced by a neutral oscillation. The generic-$m$ statement using the active harmonic
$k\equiv\pm2$ holds for $m=3$ and $m\ge5$; $m=4$ is the marked exception in
Table~\ref{tab:catalog}.

\section{Numerical methods}
\label{app:numerics}

\emph{Integration.} The model
Eqs.~\eqref{eq:modelx}--\eqref{eq:modeltheta} is integrated with a fixed-step
fourth-order Runge--Kutta scheme, vectorized over the $N$ units, with
step $dt=0.05$ (reduced to $0.01$ for the Hamiltonian-line tests, where $H$ has
absolute range $\sim10^{-12}$ over the run); phases are taken modulo $2\pi$ on output.
Trajectories are run to $t=120$ (longer near boundaries), and we treat a state as
reached when its order-parameter diagnostics are stationary to $<10^{-4}$ over the
final tenth of the run.

\emph{Classification.} States are labeled from the rainbow order parameters and the
Daido moments: sync $r,s>1-\delta$; phase wave / cluster $s>1-\delta$, $r<\delta$
(separated by the cluster moment $X_m=|\langle e^{imx}\rangle|$); async / $\mathcal M$
$r,s<\delta$; $q$-twist $w_q=|\langle e^{i(qx-\theta)}\rangle|>w_c$ on top of
$r,s<\delta$. We use $\delta=0.1$ and verified that varying $\delta$ and the moment
thresholds by factors of $2$--$10$ does not change any qualitative conclusion.

\emph{Spectral checks.} For each family we formed the full $2N\times2N$ Jacobian both
analytically (Eq.~\eqref{eq:Mblock}) and by central finite differences of the
microscopic right-hand side, and compared eigenvalues; the maximum entry and spectrum
errors were $\lesssim10^{-9}$ across sync, phase wave, spatial/phase sync,
$q$-twists ($q=2,\dots,10$) and $m$-clusters ($m=2,\dots,12$), over the four
opposite-sign coupling sub-quadrants and the same-sign quadrants.

\emph{Local-robustness experiments.} For each hidden state we kicked the exact
configuration with independent per-coordinate Gaussian noise of amplitude
$\varepsilon$, integrated, and recorded the relaxed labeling order parameter and
whether the labeled state returned. The threshold $\varepsilon^\ast$ is the crossing
of the relaxed order parameter through $w_c$ ($X_c$); the collapse plots use
$u=\varepsilon\sqrt{q^2+1}$ and $u=\varepsilon m$. We repeated this for several $N$,
several $q,m$, and several thresholds to confirm the $-1$ exponent, the
threshold-only dependence of the prefactor, and the onset of the finite-$N$ floor.

\emph{Far-kick geometry.} For the boxy/octopus test we grew $\varepsilon$ from the
core scale to the random-initial-condition scale, measured the far-return probability
$P_{\rm ret}(\varepsilon)$ over many trials, and ran the same protocol on a
nearest-neighbour Kuramoto ring (with a known stable twist) for contrast, including
the re-kick fragility test of far-returning ring points.

\emph{Basin census.} Global basin stability was estimated from $10^6$
uniformly-random initial conditions per $N$, classified as above; no hidden
structured state was recovered ($S_B<10^{-6}$). Parameter values, seeds, and the
scripts that generate every figure are provided with the manuscript.

\bibliographystyle{apsrev4-2}
\bibliography{sample}

\end{document}